# Quantifying Angular Correlations between the Atomic Lattice and Superlattice of Nanocrystals Assembled with Directional Linking

Ivan. A. Zaluzhnyy,[1,2] Ruslan P. Kurta,[3] Alexander André,[4] Oleg Y. Gorobtsov,[1] Max Rose,[1] Petr Skopintsev, [1,+] Ilya Besedin,[1,2,++] Alexey V. Zozulya,[1,3] Michael Sprung,[1] Frank Schreiber,[4] Ivan A. Vartanyants,[1,2] and Marcus Scheele[4,*]

[1]Deutsches Elektronen-Synchrotron DESY, Notkestraße 85, D-22607 Hamburg, Germany

[2]National Research Nuclear University MEPhI (Moscow Engineering Physics Institute), Kashirskoe shosse 31, 115409 Moscow, Russia

[3]European XFEL GmbH, Holzkoppel 4, D-22869 Schenefeld, Germany

[4]Eberhard Karls Universität Tübingen, Geschwister-Scholl-Platz, D-72074 Tübingen, Germany

ABSTRACT

We show that the combination of X-ray scattering with a nanofocused beam and X-ray cross correlation analysis is an efficient means for the full structural characterization of mesocrystalline nanoparticle assemblies with a single experiment. We analyze several hundred diffraction patterns of individual sample locations, i.e. individual grains, to obtain a meaningful statistical distribution of the superlattice and atomic lattice ordering. Simultaneous small- and wide-angle X-ray scattering of the same sample location allows us to determine the structure and orientation of the superlattice as well as the angular correlation of the first two Bragg peaks of the atomic lattices, their orientation with respect to the superlattice, and the average orientational misfit due to local structural disorder. This experiment is particularly advantageous for synthetic mesocrystals made by the simultaneous self-assembly of colloidal nanocrystals and surface-functionalization with conductive ligands. While the structural characterization of such materials has been challenging so far, the present method now allows correlating mesocrystalline structure with optoelectronic properties.

Mesocrystals (MC) are three-dimensional arrays of iso-oriented single-crystalline particles with an individual size between 1 – 1000 nm.[1–5] Their physical properties are largely determined by structural coherence, for which the angular correlation between their individual atomic lattices and the underlying superlattice of nanocrystals (NC) is a key ingredient.[1,2] Colloidal NCs stabilized by organic surfactants have been shown to pose excellent building blocks for the design of synthetic MCs with tailored structural properties which are conveniently obtained by self-assembly of NCs from solution on a solid or liquid substrate by exploiting ligand-ligand interactions.[6–25] Typically, the utilized ligands consist of wide-gap, bulky hydrocarbons which render the MCs insulating.[26–33] MCs obtained in this way exhibit average grain sizes of ~150 µm$^2$, which enables a detailed characterization by electron and/or X-ray microscopy.[34] Since the optoelectronic properties of PbS NC ensembles bear many opportunities for applications in solar cells or photodetectors, a number of ligand exchange procedures with small organic or inorganic molecules as well as single atom passivation strategies have been developed, all of which greatly increase the carrier mobilities within the SL of NCs.[28,33,35–44] Due to the short interparticle spacing imposed by these ligands, structural coherence is mostly lost in such superlattices, but in rare cases it has been demonstrated that significant long-range order and even mesocrystallinity can be preserved.[25,35,45] However, a persisting problem of these protocols is that they are prone to introduce defects in the superlattice structure with some degree of granularity and significantly smaller grain sizes, which poses difficulties in determining the angular correlation with a meaningful statistical distribution.[11,17,24,41,46,47] Using a conductive MC on the basis of PbS NCs,[35] we show how X-ray cross correlation analysis (XCCA)[48–51] in conjunction with a nanofocused X-ray beam can address this problem. Such a method should facilitate the application of synthetic conductive MCs with strong angular correlation for thermoelectrics, spintronics, (magneto-)electronics and optics.[52–57]

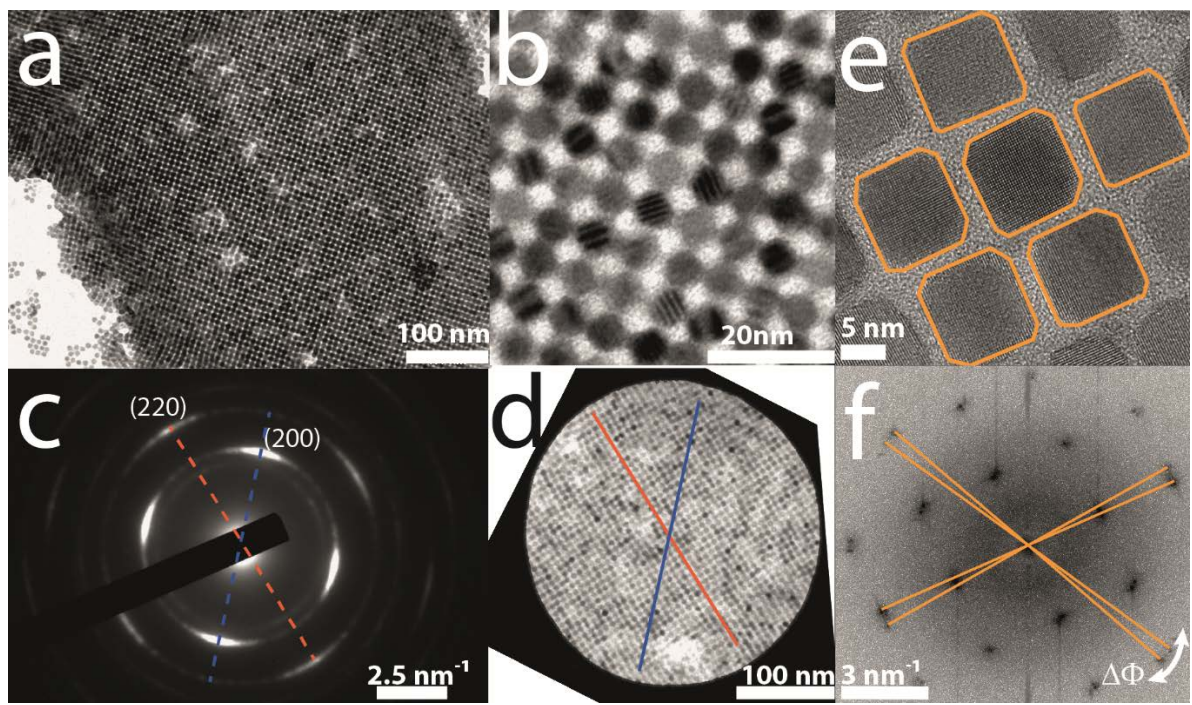

**Figure 1**. Electron micrographs of typical MCs, obtained with 6.2 nm PbS NCs cross-linked with tetrathiafulvalenedicarboxylate (TTFDA). **a**) Low-magnification and **b**) high-magnification micrograph of a typical grain showing the high degree of order in the superlattice. **c**) Electron diffraction from the selected area displayed in **d**) red and blue lines indicate mutual directions in the AL (dashed) and SL (solid). **e**) high-resolution micrograph of a monolayer of a similar sample (10.7 nm NCs) prepared under identical conditions. Boundaries of individual nanocrystals are highlighted in orange to emphasize that their atomic lattices are not precisely parallel. **f**) its corresponding Fast Fourier Transform (FFT). The split peaks in the FFT, exhibiting an orientational disorder ($\Delta\Phi$) of approx. 5 °, further support this.

**Fig. 1** displays a conductive MC of PbS NCs ligand-exchanged with the organic semiconductor tetrathiafulvalenedicarboxylate TTFDA.[35] While **Figs. 1a-b** demonstrate the typical dimensions of an individual MC grain (1-10 µm$^2$) and the high degree of order within the grain, a comparison of the electron diffraction pattern (**Fig. 1c**) with the corresponding real-space image (**Fig. 1d**) illustrates the angular correlation (blue and red lines) between the atomic lattice (AL) and the superlattice (SL). Imaging of individual NCs (**Fig. 1e** and **Fig. 1f**) reveals a small degree of misalignment between neighboring ALs, which gives rise to broadening of the diffraction peaks in **Fig. 1c**. On this highly local scale with poor statistics, the orientational disorder - represented by an angle $\Delta\Phi$ - is $\leq 5°$ (**Fig. 1f**). This angle conveys an important information about the degree of directional linking between the NCs in an SL exerted by their ligand shells. While the distribution in size and shape of the NCs is a source for inherent misalignment in such MCs, it has been shown that differences in the ligand shell can also dramatically alter $\Delta\Phi$.[11] Quantifying $\Delta\Phi$ for individual MC grains with good statistics is therefore important and will be addressed in the following by nano-diffraction.

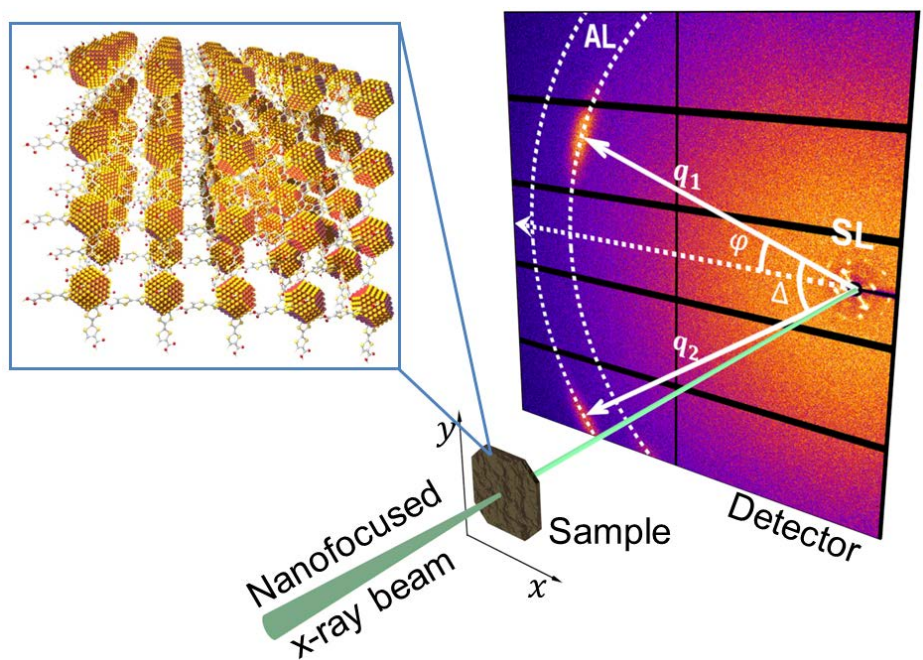

**Figure 2**. **Scheme of the diffraction experiment**. The sample is scanned by a nanofocused X-ray beam with the size of 400 nm by 400 nm in transverse direction. The detector is positioned

downstream from the sample in transmission geometry, and is shifted from the optical axis of the beam to measure simultaneously SL and AL peaks. The angle φ is measured with respect to the horizontal axis; the positive direction is counterclockwise, $q_1$ and $q_2$ are the magnitudes of the momentum transfer vectors, and $-\pi < \Delta \leq \pi$ is an angular variable. (inset) Details of the structure of the SL consisted of PbS NCs coated with TTFDA.

Due to the granularity in the sample, any macroscopic imaging technique with a meaningful statistical description of the whole sample will necessarily return the average orientation of all SLs and ALs.[4,47] For the given example, which is representative for state-of-the-art artificial, conductive MCs, this renders a determination of the angular correlation and orientational disorder within each MC grain practically impossible. To alleviate this problem, we measured X-ray nano-diffraction of an MC obtained with the same NCs under the same conditions as those displayed in **Figs. 1a-d.** A nanofocused beam with a footprint of about 400 x 400 nm$^2$, which is smaller than a typical MC domain, was used. We utilized PbS NCs obtained by wet-chemical procedures[58] since their size-distribution is relatively narrow (~5 %) (see Methods for sample preparation details), and TTFDA was chosen as a conductive linker because tetrathiafulvalene derivatives have been shown to invoke field-effect mobilities on the order of $10^{-4} – 10^{-3}$ cm$^2$/Vs in PbS ensembles.[35,44] However, we emphasize that the analysis described in this Letter is not limited to specific NC superlattices and is generally applicable to elucidate angular correlations in MCs. The experiment schematically shown in **Fig. 2** (see Methods for experimental details) gave us a remarkable opportunity to simultaneously observe scattering from two distinct length scales (AL and SL) in a single image. This allowed us to determine the structure and orientation of the SL, as well as angular correlations between AL and SL within the same MC grain.

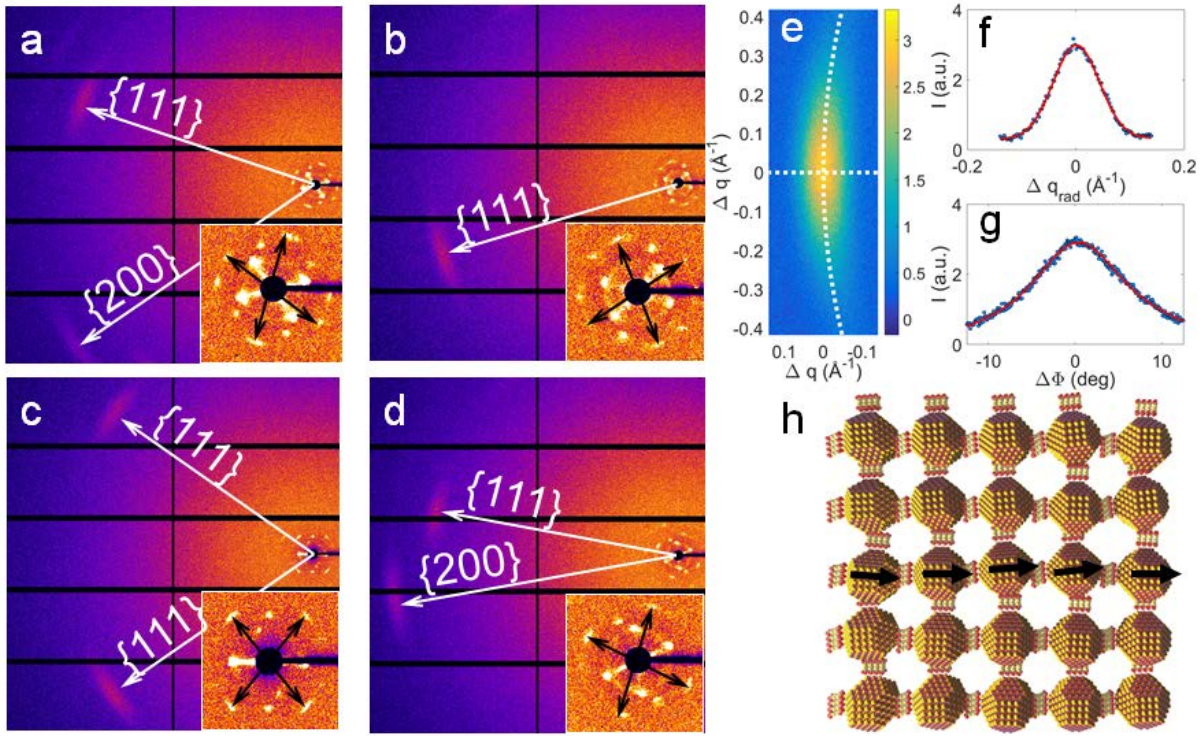

**Figure 3: Examples of measured diffraction patterns from MCs. a-d)** Diffraction patterns measured at different positions of the sample. One can see SAXS scattering from the SL and Bragg peaks, corresponding to {111} and {200} reflections of the PbS AL. The insets display enlarged SAXS regions with the scattering signal from the SL. White arrows point to the Bragg reflections from the PbS AL and black arrows show the diffraction peaks at $q_4^{SL} = 1.72$ nm$^{-1}$ from the SL. **e)** Two-dimensional shape of the 111 diffraction peak of the AL averaged over 412 diffraction patterns. White lines represent cross sections in radial and azimuthal directions. **f-g)** Radial and azimuthal cross section through the center of the peak, respectively. Points are experimental data and red lines are Gaussian fits. **h)** Idealized schematics of a PbS NC SL with a simulated orientational disorder of max. 10°. Black arrows indicate the same crystallographic direction in individual ALs. For clarity, the ligand spheres are sketched by small molecular stacks.

Typical diffraction patterns measured at different positions on the sample are shown in **Fig. 3**, displaying up to four different orders of SL peaks in the small angle scattering (SAXS) region, with the momentum transfer values $q_1^{SL} = 0.98$ nm$^{-1}$, $q_2^{SL} = 1.08$ nm$^{-1}$, $q_3^{SL} = 1.34$ nm$^{-1}$ and $q_4^{SL} = 1.72$ nm$^{-1}$ (see Supplementary **Fig. S1** for details). We also often observed wide angle X-ray scattering (WAXS) from the {111} and {200} planes of the AL at $q_{111}^{AL} = 18.3$ nm$^{-1}$ and $q_{200}^{AL} = 21.2$ nm$^{-1}$, respectively. At some positions of the sample, we observed two AL {111} peaks (see **Fig. 3c**). We scanned a sample area of 13.6 x 20 μm$^2$ in steps of 400 nm and recorded 1785 individual diffraction patterns, 412 of which displayed at least one well-resolved 111$_{AL}$ diffraction peak. To obtain a statistical description of ΔΦ, we averaged the signal of all 111$_{AL}$ diffraction peaks of the AL (**Fig. 3e**), which allowed us to extract the radial and azimuthal cross section of the averaged diffraction peak shown in **Fig. 3f** and **3g**, respectively. We find that both cross sections are well fit by Gaussian functions and that the peak is significantly broader in azimuthal than in radial direction. Attributing this additional azimuthal broadening to the orientational disorder of NCs discussed above, we obtained a value of ΔΦ ~ 10°. This - in contrast to **Fig. 1 -** represents the orientational disorder within a typical MC domain averaged over a macroscopic portion of the sample. Our result is illustrated in **Fig. 3h,** depicting an SL of PbS NCs with a simulated ΔΦ of 10°. Such rather large angular disorder may be rationalized as originating to a significant extent from the hybrid nature of the material, consisting of "hard" NCs and "soft" ligand spheres, which act as directional linkers between the NCs with some structural flexibility. Note that the real shape and binding mode of the ligand sphere may be more complex than that displayed in **Fig. 3h**.

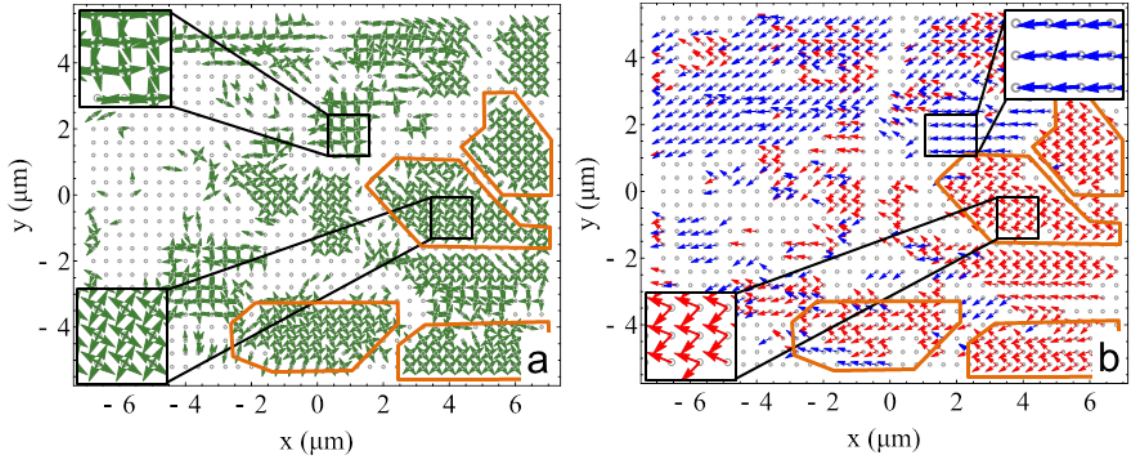

**Figure 4: Domain structure of the sample.** Spatially resolved maps of angular positions of the diffraction peaks of the SL at $q_4^{SL}$ **(a)** and PbS AL **(b)**. Red arrows represent the angular position of {111} Bragg reflection at $q_{111}^{AL}$, and blue arrows correspond to the angular position of {200} Bragg reflection at $q_{200}^{AL}$. Some domains are marked with orange lines. Two areas of the maps are enlarged in the insets for better visibility.

Furthermore, the large number of diffraction patterns allowed us to construct a spatially resolved map of the angular orientations of individual MC grains (**Fig. 4**). It is apparent from **Fig. 4** that the sample is a granular MC with typical domain sizes of about 6-8 μm$^2$ (corresponding to at least 2·10$^6$ individual PbS particles in one domain), with well-resolved borders of width of about 0.5 μm between the domains with different orientations. The predominant orientation of the AL perpendicular to the sample surface is [110]$_{AL}$, deduced from the angle of 70.5° between two {111} reflections (the AL of PbS is fcc). For each position on the sample, we also determined a mismatch angle ψ displaying the azimuthal misorientation between SL and AL (see

Supplementary **Fig. S3** for details). The mean value of this angle was ψ = (0.1±2.5)°, emphasizing that the angular correlation between SL and AL is very robust.

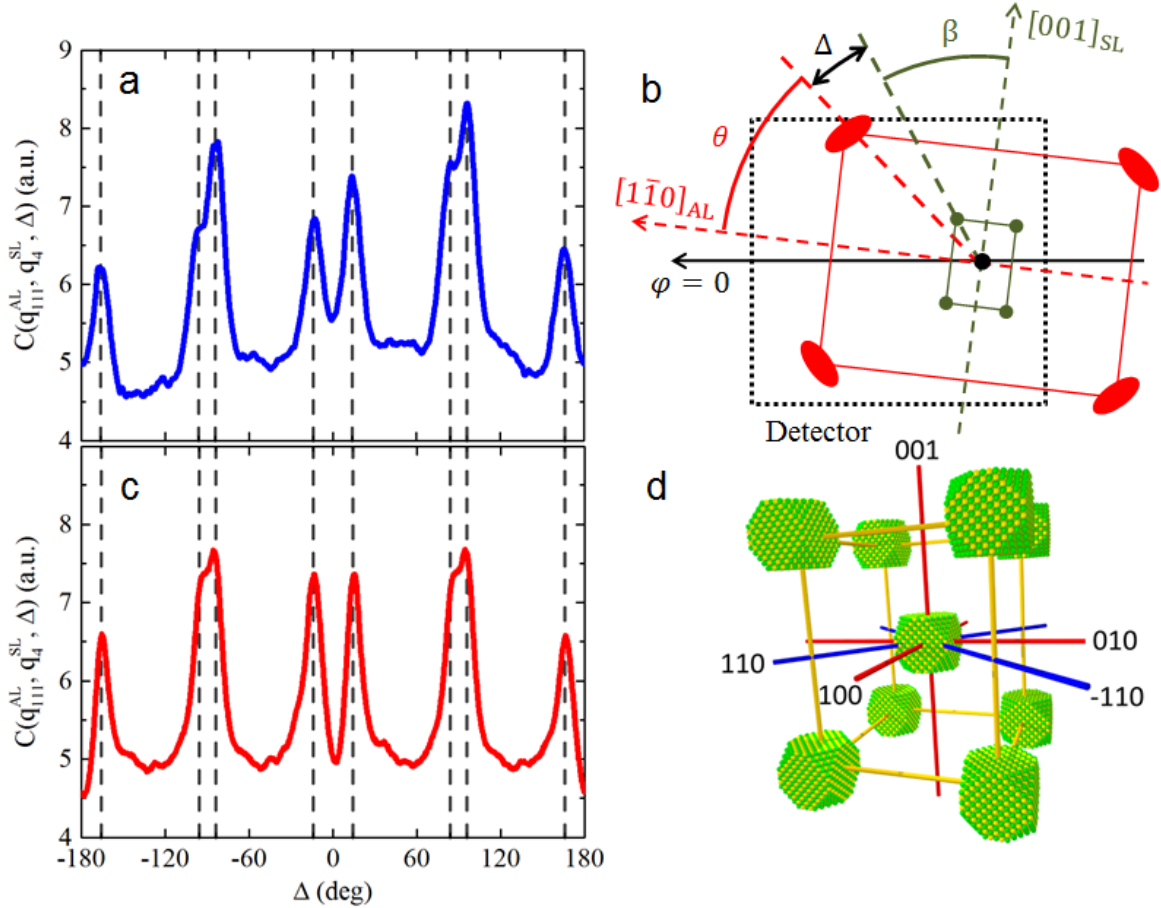

**Figure 5: Angular correlations between AL and SL. a)** Average CCF calculated from the experimental data. **b)** Schematic representation of AL and SL in reciprocal space. Red ellipsoids represent $\{111\}_{AL}$ reflections from the AL oriented in such a way that the $[110]_{AL}$ direction is perpendicular to the sample surface. The angle $\theta = \arctan(1/\sqrt{2}) \approx 35.26°$ between $[1\bar{1}0]_{AL}$ and $[1\bar{1}1]_{AL}$ directions is fixed for the fcc lattice of PbS. Green circles represent $\{112\}_{SL}$ reflections from the SL oriented in such a way that the $[110]_{SL}$ direction is perpendicular to the sample surface. The value of the angle $\beta$ between $[001]_{SL}$ and $[1\bar{1}2]_{SL}$ directions for the bct structure of the SL depends on the tetragonal distortion. To illustrate the meaning of $\Delta$, the first

angular correlation (for which $\Delta \approx 14°$) is shown. The area of reciprocal space covered by the detector is marked by a dashed rectangle. **c)** Model CCF obtained assuming that the SL has tetragonal distortion $c/a = 1.22$. **d)** Schematic of the MC unit cell displaying the angular correlation between AL and SL; collinear axes are indicated in red (<100> directions) or blue (<110> directions).

In order to determine the angular correlations and unit cell parameters within each domain quantitatively, we applied an XCCA approach (see Methods for XCCA details).[48–51,59–62] This method is based on the evaluation of a two-point angular cross-correlation function (CCF) that can be calculated for each diffraction profile as[48,49]

$$C(q_1, q_2, \Delta) = \langle I(q_1, \varphi) I(q_2, \varphi + \Delta) \rangle_\varphi. \quad (1)$$

The scattered intensity $I(q, \varphi)$ is expressed in polar coordinates and $\langle ... \rangle_\varphi$ denotes azimuthal averaging (see **Fig. 2**). The CCF has a clear geometrical meaning: it shows how well two intensity profiles, $I(q_1, \varphi)$ and $I(q_2, \varphi)$, azimuthally align with each other. Here, we determined the CCF $\langle C(q_{111}^{AL}, q_4^{SL}, \Delta) \rangle$ averaged over 412 diffraction patterns, for which scattering was observed both at $q_{111}^{AL}$ and $q_4^{SL}$. The result of this analysis is shown in **Fig. 5a.** We observed eight correlation peaks at $\Delta = \pm 14°, \pm 84°, \pm 96°, \pm 166°$, representing the relative angles between $q_{111}^{AL}$ and $q_4^{SL}$. In the following, we will show that the average CCF characterizes the structure of the MC, including the angular correlations between AL and SL. In the present case, the PbS AL is fcc, its preferred orientation is $[110]_{AL}$ perpendicular to the film's surface, and geometric analysis of typical diffraction patterns suggests a bcc lattice for SL with predominant $[110]_{SL}$ orientation parallel to the same orientation of the AL. The symmetric profile of the CCF with respect to $\Delta=0°$ also indicates collinearity of the $[001]_{SL}$ and $[001]_{AL}$ directions (see Methods for details). However, one can show that such two superimposed fcc and bcc lattices would be characterized by an average CCF with only six peaks at $\Delta =\pm19.4°, \pm90°, \pm160.6°$, inconsistent with the eight peaks observed in our experiment. We found that the experimental CCF can be reproduced if one considers a tetragonal distortion of the bcc lattice (to a body-centered tetragonal (bct) structure) with the unit cell parameters $a = b \neq c$. For a bct lattice, the smallest angle $2\beta = 2\arctan\left(c/a\sqrt{2}\right)$ between two $[112]_{SL}$ directions is determined by the tetragonal distortion $c/a$ and can be obtained by means of the CCF (see **Fig. 5b**). In this case, the CCF peaks appear at $\Delta = \pm(90° \pm \theta \pm \beta)$, where $\theta = 35.3°$ is the angle between the $[110]_{AL}$ and $[111]_{AL}$ directions of the fcc AL. Therefore, one can expect eight permutations for $\Delta$ in qualitative agreement with the CCF in **Fig. 5a.** This way, a comparison with our experimental CCF in **Fig. 5a** gave us value of $\beta \approx 40.7°$ and correspondingly a tetragonal distortion with $c/a \approx 1.22$.

To verify these findings, we modelled a CCF under the assumption of an fcc AL and bct SL with $c/a = 1.22$ and the alignment as detailed in **Fig. 5b** (see Supplementary Materials for details of CCF modeling). The resulting model CCF shown in **Fig. 5c** is in excellent agreement with the experimental CCF (**Fig. 5a**) in terms of the positions and magnitudes of all eight peaks. We did not observe any additional broadening of the CCF peaks compared to the angular width of the diffraction peaks, which is further evidence of a strong angular correlation between the SL and AL. With respect to our main goal to elucidate the structure of the granular MC depicted in **Fig. 1**, we find that the combination of nano-diffraction and XCCA demonstrated here is particularly powerful for the characterization of MCs with different symmetries of the AL and SL. In such a case, an unambiguous determination of the angular correlation requires the analysis of many grains in different orientations, since the number of collinear axes shared between the SL and AL is much smaller than for MCs where AL and SL have the same symmetry. This is illustrated in **Fig. 5d** for the present case, which is directly derived from our XCCA, and features a total of five collinear axes shared between both lattices: all three <100> axes as well as the family of <110>-directions. Note that there is no collinearity along the [011] direction or the family of <111> directions. The latter result is particularly noteworthy since the <111>$_{SL}$ direction is the nearest-neighbor distance in a body-centered structure, and one may intuitively have expected an iso-orientation, which is not observed. The fact that the shorter cubic directions of the SL ("a") exhibit collinearity with the AL, and the longer cubic direction ("c") does not, supports the view that ligand-ligand interactions from specific facets of neighboring NCs are responsible for the often observed tetragonal distortion in PbS MCs.[24,63]

We anticipate that the progressive exploration of synthetic MCs with increasing complexity, e.g, binary NC superlattices, nanorod assemblies, honeycomb lattices, *etc.* will benefit strongly from the present study. As we have demonstrated, a single experiment is sufficient to quantify the structure of the superlattice, its angular correlation with the atomic lattices, the average orientational disorder between atomic lattices and a meaningful statistical distribution of these parameters as well as the length scale of the domains and their boundaries. This should greatly facilitate the understanding of structure-property relationships in MCs.

ASSOCIATED CONTENT

**Supporting Information**. The following files are available free of charge: Description of the sample preparation, details of the X-ray scattering experiment, the X-ray cross-correlation analysis, the evaluation of the model cross-correlation function and a comment on the effect of orientational disorder between atomic lattices. **Figure S1**: Analysis of the angular averaged signal in SAXS region. **Figure S2**: Evaluation of the angle 2β between two {112}SL peaks for the bct structure. **Figure S3**: Spatially resolved map of misorientation between the superlattice and atomic lattice.

AUTHOR INFORMATION

**Corresponding Author**

* marcus.scheele@uni-tuebingen.de

**Present Addresses**

[+] Laboratory for Biomolecular Research, Paul Scherrer Institute, Villigen 5232, Switzerland

[++] National University of Science and Technology MISIS, Leninsky prospect 4, 119049 Moscow, Russia

**Author Contributions**

The manuscript was written through contributions of all authors. All authors have given approval to the final version of the manuscript.

**Funding Sources**

This work was supported by the Virtual Institute VH-VI-403 of the Helmholtz Association as well as the DFG (SCHE1905/3; SCHE1905/4 and SCHR700/25).

ACKNOWLEDGMENT

We acknowledge support by E. Weckert. We thank T. Salditt for providing nano-focusing instrument (GINIX) support at P10 beamline. High-resolution TEM imaging (Fig. 1e) and the synthesis of TTFDA were performed as user projects at the Molecular Foundry, which was supported by the Office of Science, Office of Basic Energy Sciences, of the U.S. Department of Energy under Contract No. DE-AC02-05CH11231.

# Supporting Information

**Sample preparation**

The two batches of PbS nanocrystals utilized in this study with diameters of 6.2 nm (for all X-ray experiments and the electron micrographs in **Figs. 1a-d** and 10.7 nm (for the electron micrographs in **Figs. 1e-f**) were synthesized following Weidman *et al.*[1] Neutral tetrathiafulvalene dicarboxylic acid was synthesized, deprotonated to TTFDA and used as a directional linker during a solid/air assembly of PbS nanocrystals into superlattices as previously described.[2] For electron microscopy, copper grids coated with a thin amorphous carbon film were applied as substrates. For X-ray experiments, we utilized 5 mm x 5mm Si frames with a 500 µm x 500 µm window consisting of a 50 nm thick $Si_3N_4$ membrane (PLANO) as substrates.

**X-ray diffraction experiment**

The X-ray diffraction experiment was conducted at the Coherence Beamline P10 of the PETRA III synchrotron source at DESY. The nanodiffraction endstation GINIX was used to focus an X-ray beam with energy $E = 13.8$ keV ($\lambda = 0.898$ Å) down to 400 x 400 nm$^2$ size with KB-mirrors.[3] The depth of the X-ray focus was about 0.5 mm. The sample was positioned perpendicular to the incoming X-ray beam, and an area of 13.6 x 20 µm$^2$ was scanned to analyze the spatial variations of the samples' structure. Within this scanning region, 1785 diffraction patterns were collected on the 34 x 50 raster grid with a 400 nm step size in both directions. Each diffraction pattern was collected with an exposure time of 0.5 s to prevent radiation damage, which was confirmed by repeating the scanning procedure several times on the same position of the sample. A two-dimensional detector Pilatus 1M (981x1043 pixels of 172x172 μm² size) was positioned downstream at a distance of 46 cm from the sample and shifted approximately by 8 cm to the side (**Fig. 2**). In such geometry, we were able to detect the scattering signal from the SL as well as from PbS AL simultaneously. Only a part of reciprocal space in wide angle scattering was accessible with the detector. The detector resolution of SAXS was about 0.03 nm$^{-1}$ and in the region of WAXS – approximately 0.02 nm$^{-1}$. The measured signal was corrected for background scattering.

**X-ray cross-correlation analysis**

The CCFs $C(q_1, q_2, \Delta)$ were calculated according to the equation

$$C(q_1, q_2, \Delta) = \frac{1}{2\pi} \int_{-\pi}^{\pi} I(q_1, \varphi) I(q_2, \varphi + \Delta) d\varphi,$$

where $I(q, \varphi)$ is expressed in polar coordinates on the detector plane, $q_1$ and $q_2$ are two magnitudes of the momentum transfer vectors, $\varphi$ is an angular coordinate around a diffraction ring, $-\pi < \Delta \leq \pi$ is an angular variable (**Fig. 2**). To obtain statistically meaningful data, CCFs were averaged over a large number of measured diffraction patterns

$$\langle C(q_1, q_2, \Delta) \rangle_M = \frac{1}{M} \sum_{i=1}^{M} C^i(q_1, q_2, \Delta).$$

Here index $i$ enumerates diffraction patterns, $M$ is total number of measured diffraction patterns and $\langle ... \rangle_M$ denotes ensemble averaging.[4]

In the present work, the scattered signal in the WAXS region was limited by the detector size. In this case, one can still evaluate two-point CCF $C(q_1, q_2, \Delta)$ by applying a mask $W(q, \varphi)$, which is equal to unity in the regions where the scattering signal was measured and to zero outside of the detector. Using such a mask allows one to consider only a pair of points in the CCF that both lie within the detector area. In this case, the CCF should be evaluated as

$$C(q_1, q_2, \Delta) = \frac{\int_{-\pi}^{\pi} I(q_1, \varphi) W(q_1, \varphi) I(q_2, \varphi + \Delta) W(q_2, \varphi + \Delta) d\varphi}{\int_{-\pi}^{\pi} W(q_1, \varphi) W(q_2, \varphi + \Delta) d\varphi}.$$

In our work, the signal in the SAXS region was recorded by the detector completely, so $W(q_4^{SL}, \varphi) = 1$, except for the region shadowed by the beamstop holder, where $W(q_4^{SL}, \varphi) = 0$. In the WAXS region the mask $W(q_{111}^{AL}, \varphi)$ was set to zero outside the detector (for $\varphi \lesssim -45°$ and $\varphi \gtrsim 45°$) and in the region of detector gaps. For all other points, $W(q_{111}^{AL}, \varphi) = 1$.

**Analysis of the angular averaged signal in SAXS region**

In order to determine the structure of the SL we performed analysis of the angular averaged scattered signal in the SAXS region. After subtraction of a measured background from the diffraction data and integration over the azimuthal variable $\varphi$, we obtained a q-dependence of the SAXS signal $I(q)$, which is shown in **Fig. S1a**. We observe four scattering peaks from the SL at $q_1^{SL} = 0.98$ nm$^{-1}$, $q_2^{SL} = 1.08$ nm$^{-1}$, $q_3^{SL} = 1.34$ nm$^{-1}$ and $q_4^{SL} = 1.72$ nm$^{-1}$. The first two peaks are not resolved completely, if the signal averaged over all measured diffraction patterns is considered. However, on individual diffraction profiles two peaks at $q_1^{SL}$ and $q_2^{SL}$ are well distinguished. We attribute these peaks to the {011}, {110}, {002} and {112} families of reflections of bcc lattice with tetragonal distortion. If we fix the value of the tetragonal distortion obtained from XCCA ($c/a \approx 1.22$), the best fitting of the diffraction peak positions was obtained for a bct structure with unit cell parameters $a = b = 7.9 \pm 0.4$ nm and $c = 9.7 \pm 0.4$ nm (for such a lattice $q_1^{SL} = 1.03$ nm$^{-1}$, $q_2^{SL} = 1.13$ nm$^{-1}$, $q_3^{SL} = 1.30$ nm$^{-1}$ and $q_4^{SL} = 1.72$ nm$^{-1}$)

Due to the inhomogeneity of the sample and imperfections of the SL, the determination of SL unit cell parameters becomes challenging. In **Fig. S1b** the q-positions of the {011}, {110}, {002} and {112} reflections are shown as a function of the diffraction pattern number. It is clearly visible that these values fluctuate depending on the position on the sample, which means that the structure of the SL is slightly position-dependent.

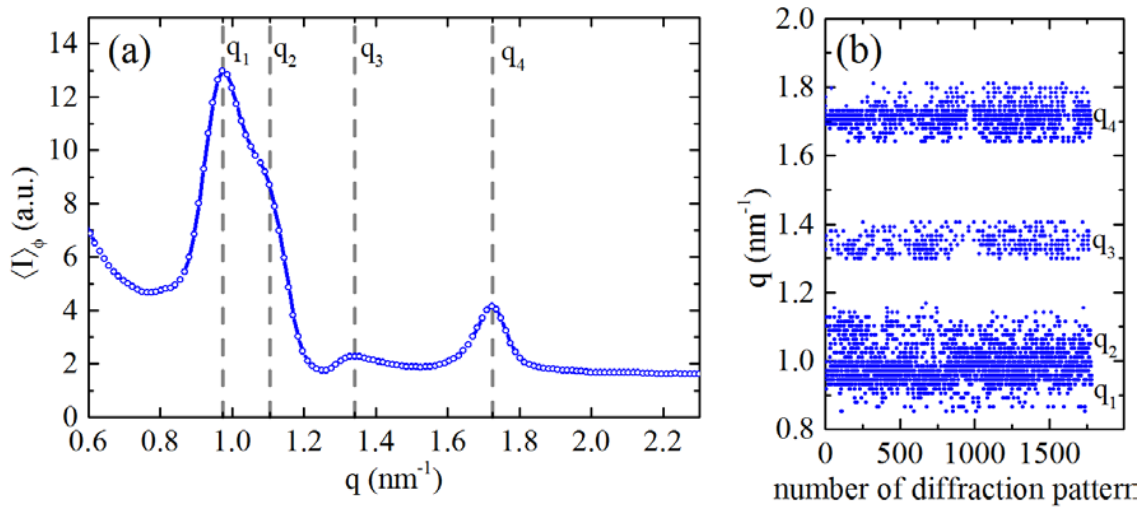

***Figure S1.*** *a) The q-dependence of the angular averaged intensity $\langle I(q, \varphi)\rangle_\varphi$. b) q-position of the scattering peaks for different individual diffraction patterns*

**Evaluation of angle $2\beta$ between two {112}$_{SL}$ peaks for the bct structure**

The unit cell of the bct structure is based on three orthogonal vectors $\boldsymbol{a}$, $\boldsymbol{b}$ and $\boldsymbol{c}$ with moduli $|\boldsymbol{a}| = |\boldsymbol{b}| = a$ and $|\boldsymbol{c}| = c$ (see **Fig. S2a**). Corresponding vectors of the reciprocal lattice are also orthogonal, so the Gram matrix for the reciprocal space is

$$G = \begin{pmatrix} 2\pi/a^2 & 0 & 0 \\ 0 & 2\pi/a^2 & 0 \\ 0 & 0 & 2\pi/c^2 \end{pmatrix}. \quad (1)$$

A diffraction pattern from a bct lattice oriented in such a way that the [110] direction is parallel to the incident X-ray beam is shown in **Fig. S2b**. The angle $2\beta$ between two Bragg peaks from {112} reflections can be evaluated as

$$\cos 2\beta = \frac{u^T G v}{\sqrt{u^T G u}\sqrt{v^T G v}}, \quad (2)$$

where $u^T = (1\ -1\ 2)$ and $v^T = (-1\ 1\ 2)$ are the indexes of two reflections and T denotes transposition. Substituting the Gram matrix (1) into Eq. (2) and solving the obtained equation for the angle $\beta$ yields

$$\tan\beta = \frac{c}{a\sqrt{2}}. \quad (3)$$

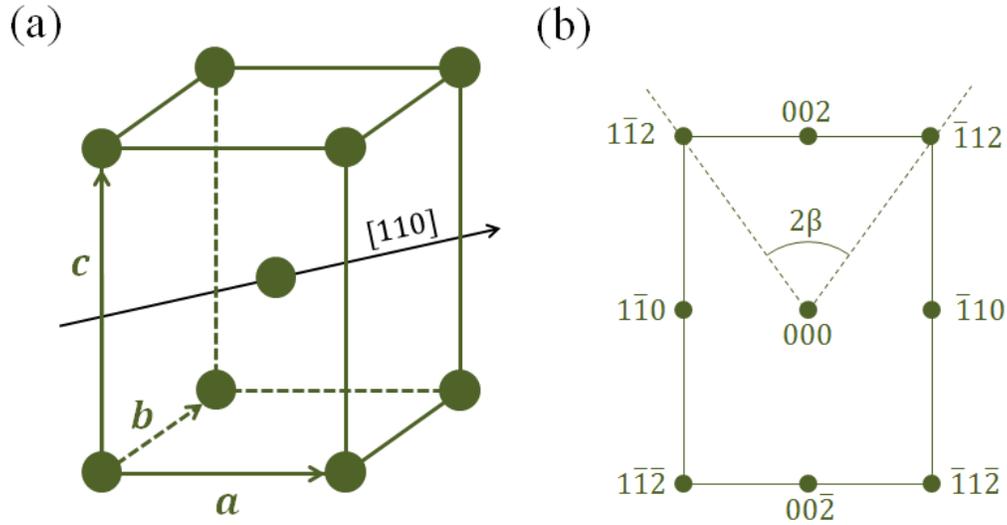

*Figure S2. a) Unit cell of the bct structure. b) Positions of the Bragg peaks from the bct structure if [110] direction is parallel to the beam.*

**Spatially resolved map of misorientation between the superlattice and atomic lattice**

At many positions on the sample, one can unambiguously determine both orientations of SL and AL (see spatially resolved maps in **Figs. 4a-b**). For these positions, we can directly calculate a mismatch between the angular orientation of SL and AL. Let us assume that the directions $[110]_{AL}$ and $[110]_{SL}$ coincide and they are perpendicular to the sample surface. In this case, both SL and AL can rotate around this axis. Thus, we can measure an angular mismatch $\psi$ between SL and AL by considering an angle between $[001]_{AL}$ and $[001]_{SL}$ directions (or equivalent angle between $[1\bar{1}0]_{AL}$ and $[1\bar{1}0]_{SL}$ directions). In case of perfect alignment this angle is equal to zero, and it can be positive or negative depending on the direction in which PbS particles are rotated with respect to SL (see **Fig. S3a**).

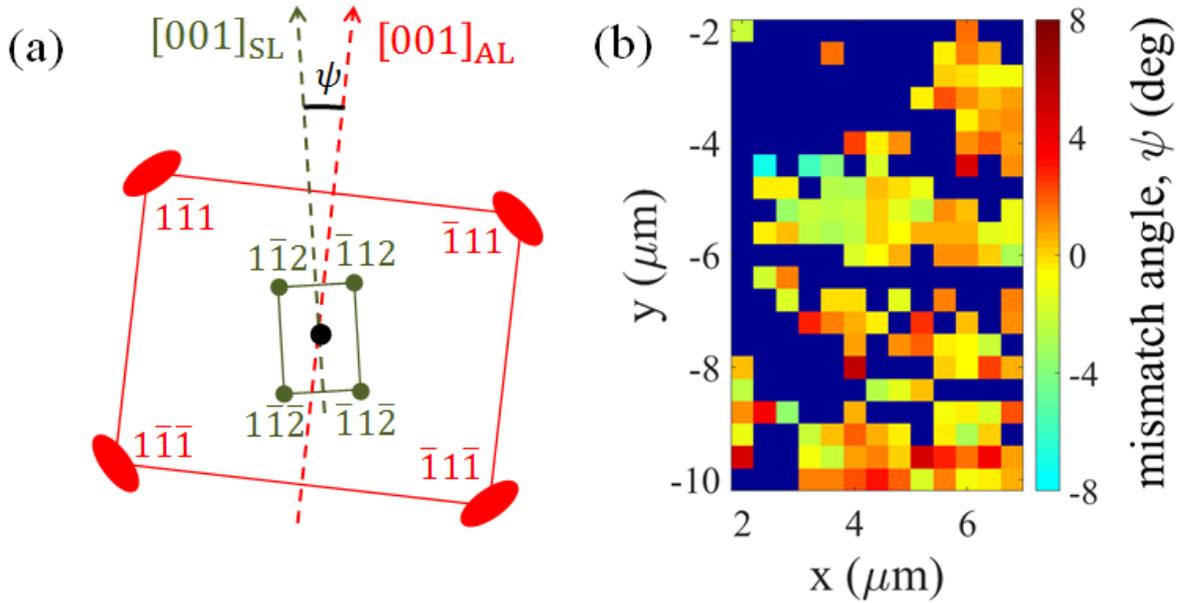

*Figure S3. a) Schematic representation of reciprocal space. Diffraction peaks from the AL are shown with red color, and diffraction peaks from the SL are shown with green color. A mismatch angle ψ between $[001]_{AL}$ and $[001]_{SL}$ directions represents misalignment between SL and AL. b) Spatially resolved map showing the angular mismatch between orientations of SL and AL. Dark blue regions correspond to points on the sample where orientation of the SL or AL, or both, could not be determined. Typically, the angular mismatch between SL and AL lies in the range from -2.5° to 2.5°.*

The spatially resolved map of the mismatch angle ψ for some region of the sample is shown in **Fig. S3b**. Different values of the angle are represented by different colors, and dark blue color corresponds to the regions, where the mismatch angle could not be obtained, due to absence of scattering signal or ambiguity in determination of SL or AL orientation. The mean value of the mismatch angle is equal to approximately 0.1°, which indicates that PbS nanoparticles can be rotated in any direction with respect to the SL (otherwise the mean value would significantly differ from zero). The standard deviation for the mismatch angle is about 2.5°, indicating that the PbS nanoparticles are aligned with respect to the SL and can only slightly be rotated around the preferred orientation within an angle of few degrees.

**On the effect of orientational disorder between atomic lattices**

If all atomic lattices were perfectly aligned with respect to each other (that is, if $\Delta\Phi$ in **Fig. 1f** and **Fig. 3h** was zero), the PbS [110] direction would be precisely parallel to the incident X-ray beam for all atomic lattices. In such a case, the Ewald sphere does not intersect with the PbS {111} Bragg reflections, even upon taking into account the finite size of the PbS nanocrystals and the associated broadening of the Bragg peaks. However because $\Delta\Phi \neq 0$, significant intersection with the Ewald sphere is possible for most nanocrystals albeit the general [110] orientation. This is why the {111} Bragg peaks are indeed frequently observed by us (see **Fig. 3a-d**).

**Evaluation of the model cross-correlation function**

For a realistic model of the experimental conditions, it is not enough to take into account the size of the detector that limits the measured scattering signal in the WAXS region. One also has to consider the fact that the sample can be oriented differently with respect to the detector. Thus, an angle δ between the $[1\bar{1}0]_{AL}$ direction of the AL and direction $\varphi = 0$ is not necessarily equal to zero (see **Fig. 5b**). Moreover, this angle can vary from one diffraction pattern to another. Due to that we calculated the model CCF as an average over all possible values of angle $\delta$

$$C_{model}(\Delta) = \frac{1}{\sqrt{2\pi\sigma_\delta^2}} \int_{-\pi}^{\pi} C(\Delta, \delta) exp\left[-\frac{1}{2}\frac{(\delta-\delta_0)^2}{\sigma_\delta^2}\right] d\delta. \quad (4)$$

Here, we assume a normal distribution of the angle $\delta$. The best fit with the experimental CCF was obtained for the standard deviation $\sigma_\delta = 0.15$ (8.6°) and mean $\delta_0 = 0.045$ (2.6°). A non-zero value of parameter $\delta_0$ indicates that the sample was misaligned with respect to the detector. The CCF $C(\Delta, \delta)$ was obtained according to the following procedure

$$C(\Delta,\delta) = \frac{\int_{-\pi}^{\pi} I_{WAXS}(\varphi,\delta) W_{WAXS}(\varphi) I_{SAXS}(\varphi+\Delta,\delta) W_{SAXS}(\varphi+\Delta) d\varphi}{\int_{-\pi}^{\pi} W_{WAXS}(\varphi) W_{SAXS}(\varphi+\Delta) d\varphi}. \quad (5)$$

Here $W_{WAXS}(\varphi)$ and $W_{SAXS}(\varphi)$ are masks in the WAXS and SAXS regions, respectively. The intensity of the WAXS peaks is modeled by a sum of Lorentzian functions

$$I_{WAXS}(\varphi,\delta) = A_{WAXS} \sum_{i=1}^{4} \frac{\gamma_{WAXS}^2}{\left(\varphi-\delta-\varphi_{WAXS}^i\right)^2 + \gamma_{WAXS}^2}, \quad (6)$$

where $A_{WAXS} = 2.3$ is a constant scaling factor, $\gamma_{WAXS} = 0.09$ (5.2°) is the half width at half maximum of a diffraction peak. If the coordinate system is introduced as shown in **Fig. 5b,** the angular positions of the $111_{AL}$ diffraction peaks in WAXS region are $\varphi_{WAXS}^i = \{-\pi+\theta, -\theta, \theta, \pi-\theta,\} \approx \{-144.74°, -35.26°, 35.26°, 144.74°\}$, where $\theta = \arctan\left(\frac{1}{\sqrt{2}}\right)$. The intensity from the bct structure along the [110] zone axis, $I_{SAXS}(\varphi)$, was modelled in the same way as a sum of four Lorentzian peaks

$$I_{SAXS}(\varphi,\delta) = A_{SAXS} \sum_{i=1}^{4} \frac{\gamma_{SAXS}^2}{\left(\varphi-\delta-\varphi_{SAXS}^i\right)^2 + \gamma_{SAXS}^2}, \quad (7)$$

where $A_{SAXS} = 17$, $\gamma_{SAXS} = 0.03$ (1.7°), $\varphi^i_{SAXS} = \{-\pi/2 - \beta, -\pi/2 + \beta, \pi/2 - \beta, \pi/2 + \beta\}$, and $\beta$ is an adjustable parameter. To simulate experimental conditions, we also added a noise to $I_{WAXS}(\varphi, \delta)$ and $I_{SAXS}(\varphi, \delta)$, modelled as a uniformly distributed signal in the range from -2 to 2. Using the values indicated above for the angular positions of the diffraction peaks in Eqs. (6) and (7), we assumed that the SL and AL are perfectly aligned with respect to each other, i.e. $\psi = 0$. Fitting the positions of the model CCF peaks to the experimentally obtained values yields $\beta = 40.7°$.

In the experiment, we were able to measure simultaneously the signal in the SAXS region for all azimuthal angles, however in the WAXS region we were restricted by the detector size and measured the scattering signal only in the angular range of approximately 90° azimuthally. To simulate the effect of finite detector size we used the following mask for the WAXS signal

$$W_{WAXS}(\varphi) = \begin{cases} 1, & -45° < \varphi < 45° \\ 0, & \text{otherwise} \end{cases}, \quad (8)$$

while $W_{SAXS}(\varphi) \equiv 1$ for $-180° \leq \varphi < 180°$. To represent the effect of the detector gaps, beamstop and shadow from the beamstop holder, we also set these points to zero in $W_{WAXS}(\varphi)$ and $W_{SAXS}(\varphi)$.